\documentclass[twocolumn,showpacs,preprintnumbers,amsmath,amssymb]{revtex4}

\usepackage{graphicx}
\usepackage{dcolumn}
\usepackage{bm}

\newtheorem{theorem}{Theorem}

\let\a=\alpha   \let\g=\gamma     \let\d=\delta  \let\e=\varepsilon
  \let\h=\eta    \let\k=\kappa   \let\l=\lambda
\let\m=\mu    \let\n=\nu            \let\p=\pi      \let\r=\rho
\let\s=\sigma         
 \let\o=\omega 
\let\G=\Gamma      \let\L=\Lambda  
           
  \let\jm=\jmath

\def\ps{\psi^{\phantom{*}}}
\def\bps{\bar\psi^{\phantom{*}}}

\def\\{\hfill\break} \let\==\equiv

\let\io=\infty 

\let\0=\noindent

\def\ra{{\big\rangle}}
\def\la{{\big\langle}}

\let\dpr=\partial

\def\tende#1{\,\vtop{\ialign{##\crcr\rightarrowfill\crcr
 \noalign{\kern-1pt\nointerlineskip}
 \hskip3.pt${\scriptstyle #1}$\hskip3.pt\crcr}}\,}
\def\otto{\,{\kern-1.truept\leftarrow\kern-5.truept\to\kern-1.truept}\,}

\def\NN{{\cal N}}

\def\der{{\rm d}}
\def\T#1{{#1_{\kern-3pt\lower7pt\hbox{$\widetilde{}$}}\kern3pt}}
\def\VVV#1{{\underline #1}_{\kern-3pt
\lower7pt\hbox{$\widetilde{}$}}\kern3pt\,}
\def\W#1{#1_{\kern-3pt\lower7.5pt\hbox{$\widetilde{}$}}\kern2pt\,}

\def\indica{\leaders \hbox to 0.5cm{\hss.\hss}\hfill}
\def\guida{\leaders\hbox to 1em{\hss.\hss}\hfill}
\mathchardef\oo= "0521

\def\pp{{\bf p}}\def\xx{{\bf x}}
\def\yy{{\bf y}}\def\kk{{\bf k}}
\def\zz{{\bf z}}

\def\oo{{\underline \omega}}

\def\qed{\raise1pt\hbox{\vrule height5pt width5pt depth0pt}}

\def\indic{\hbox{\raise-2pt \hbox{\indbf 1}}}

\def\virg{\quad,\quad}

\def\pp{{\bf p}}\def\xx{{\bf x}}
\def\yy{{\bf y}}\def\kk{{\bf k}}
\def\zz{{\bf z}}

\def\oo{{\underline \omega}}

\def\qed{\raise1pt\hbox{\vrule height5pt width5pt depth0pt}}

\def\indic{\hbox{\raise-2pt \hbox{\indbf 1}}}

\def\virg{\quad,\quad}

\begin{document}

\preprint{APS/123-QED}

\title{Non-perturbative Anomalies in $d=2$ QFT}

\author{G. Benfatto}
\author{P. Falco}
\author{V. Mastropietro}

\affiliation{Universit\`a di Roma Tor Vergata}

\begin{abstract} We present the first rigorous construction
of the QFT Thirring model, for any value of the mass, in a
functional integral approach, by proving that a set of Grassmann
integrals converges, as the cutoffs are removed, to a set of
Schwinger functions verifying the Osterwalder-Schrader axioms. The
massless limit is investigated and it is shown that the Schwinger
functions have different properties with respect to the ones of
the well known exact solution: the Ward Identities have anomalies
violating the anomaly non-renormalization property and additional
anomalies, apparently unnoticed before, are present in the closed
equation for the interacting propagator, obtained by combining a
Schwinger-Dyson equation with Ward Identities.
\end{abstract}


\maketitle

\section{Introduction}

The Thirring model has been the subject of a very intense research
in the last fifty years: it is one of the very few QFT models for
which non-perturbative informations can be obtained and it shares
with more realistic $d=4$ models, like $QED_4$, many features, as
it is apparent from a classical perturbative Feynman graph
analysis \cite{[MT]}. In the {\it massless case}, a complete set
of correlations has been obtained via an exact solution
\cite{[J],[K]}; they verify the Wightman recontruction axioms
\cite{[DFZ],[CRW]} so that a QFT corresponding to the massless
Thirring model can be constructed from them. In the {\it massive
case}, the equivalence (at a perturbative level) with the
Sine-Gordon model is known \cite{[C]} and some eigenstates by
Bethe-ansatz analysis have been found \cite{[BT]}; but a complete
rigorous construction has never been performed.

In this paper we fill this gap by considering a set of Grassmann
integrals regularized via suitable cutoffs and with a contact
current-current interaction and proving that, removing cutoffs and
for any value of the mass, they converge to a set of Schwinger
functions verifying the Osterwalder-Schrader reconstruction axioms
for Euclidean QFT \cite{[OS]}; this provides the first rigorous
construction of the massive Thirring model. Moreover, even if in
the massless case other constructions were known, we find in any
case interesting to reach a complete non-perturbative construction
of the Thirring model relying only on a functional integral
approach, which could be the only possible one at higher
dimensions or for more realistic models. We stress that our
results are {\it non-perturbative}, in the sense that the
Grassmann integrals are expressed in terms of series expansion
whose {\it convergence is proved} (that is we resum the naive
perturbation theory and we prove the convergence of the resummed
expansion). Our results are smooth in the mass and the massless
limit can be investigated; the natural question is wether or not
the correlations of the massive Thirring model, which we can
compute from a functional integral approach, pass over smoothly
into the known correlations of the exact solution in the massless
limit. Contrary to what was find in the well known paper
\cite{[GL]}, based on Zimmermann's version of BPH subtraction
scheme (the analysis was purely perturbative), we get that the
correlations in the massless limit have {\it different properties}
with respect to the exact solution. The Ward Identities (WI) have
anomalies violating the {\it anomaly non-renormalization}
property; moreover, {\it additional anomalies}, apparently
unnoticed before, are present even in the closed equation for the
interacting propagator, obtained by combining a Schwinger-Dyson
equation with the Ward Identities.

\section{The exact solution}
It is worth to recall the Johnson solution \cite{[J]} of the
massless Thirring model, describing $d=2$ massless Dirac fermions
with coupling $(\l/2)\int\jm^\m_\xx\jm^\m_\xx $, in order to
compare its properties with the results from the functional
integral approach. The solution is essentially based on a {\it
self-consistency} argument: a number of reasonable requirements on
the correlations is assumed, from which their explicit expression
can be determined. Calling $G^\m(\zz;\xx,\yy)$,
$G^{\m,5}(\zz;\xx,\yy)$ and $G(\xx,\yy)$ the truncated vacuum
expectations in the Minkowski space of the $T$-product of
$Z^{-1}\jm^\m_\zz\ps_\xx\bps_\yy$, $Z^{-1}\jm^{\m,5}_\zz
\ps_\xx\bps_\yy$ and $Z^{-1}\ps_\xx\bps_\yy$ respectively, where
$Z$ is the {\it wave function renormalization}, the first
assumption is the validity of WI of the form
\begin{eqnarray}
\dpr_\zz^\m G^\m(\zz;\xx,\yy)=
-ia\big[\d(\zz-\xx)-\d(\zz-\yy)\big]G(\xx,\yy)\label{w1}
\\
\dpr_\zz^\m G^{\m,5}(\zz;\xx,\yy)=
-i\bar a[\d(\zz-\xx)-\d(\zz-\yy)]\g^5G(\xx,\yy)\label{w2}
\end{eqnarray}
where $a,\bar a$ are parameters to be determined and
$\jm^\m_\zz,\jm^{\m,5}_\zz$ are operators formally defined, via a
point splitting procedure, from $\bar\psi\bar\g^\m\psi$ and
$\bar\psi\bar\g^\m\bar\g^5\psi$; $\bar\g^\m,\bar\g^5$ are the
Minkowski gamma matrices. Assuming also the validity of a {\it
Schwinger-Dyson} equation and using the WI for $G^\m$, a {\it
closed equation} for the two point Schwinger function is obtained,
which reads, if $\widehat G(\kk)$ denotes the Fourier transform of
$G(\xx,\yy)$:
\begin{equation}
\widehat G(\kk)=\widehat g(\kk)
\left[{1\over Z} -\l (a-\bar a)
\int\!{\der^2\pp\over (2\p)^2}\
 {\widehat G(\kk-\pp)\over \not\!\pp}
\right]\;.
\end{equation}
This equation can be solved (at a very formal level, as one has to
take a vanishing wave function renormalization) and $G(\xx,0)$ is
found to be equal to
$i(\bar\g^\m\partial_\m)^{-1}({|\xx|/x_0})^{-\h}$, where $x_0$ is
an arbitrary constant with the dimension of a length and $\h$ is a
critical index related to the coefficients $a,\bar a$ by
\begin{equation}
\h={\l\over 2\p} (a-\bar a)\;.\label{xxx}
\end{equation}
Finally, by a self-consistency argument
involving also the four point Schwinger
function,  the explicit values for
the anomalies were found:
\begin{equation}
a^{-1}=1-{\l\over2\pi}\quad, \quad\bar a^{-1}=1+ {\l\over2\pi}
\;.\label{xx11}
\end{equation}
The above equation is particularly significant, as it says the
anomalies do not receive contributions from higher orders; this
property is called {\it anomaly non-renormalization} and it holds,
as a statement valid at all order in perturbation theory, in
realistic models like $QED_4$ (it is the content of the well known
{\it Adler-Bardeen theorem} \cite{[AB]}). The order by order
analysis of \cite{[AB]} can be adapted to the Thirring model
\cite{[GR]} and (\ref{xx11}) is indeed obtained; hence the
validity of (\ref{xx11}) has been considered {\it
non-perturbative} verification of the {\it Adler-Bardeen theorem}
applied to the Thirring model. However, it should be noticed that
the regularization and the assumptions in the exact solution or in
the perturbative (or functional integral) approach are different,
hence there is no guarantee that the same results should be found
in the two approaches, see for instance \cite{[AF]}.

\section{Non-perturbative Renormalization}
Our starting point is the generating functional of the truncated
Schwinger functions, $W_{\k,K}(A,\h)$, defined so that
$e^{W_{\k,K}(A,\h)}$ is the following {\it Grassmann integral}:
\begin{equation}
{1\over\NN}\int\! P_{\k,K}(d\psi)e^{\int\!d\xx\left[-{\l\over 2}
\jm^\m_\xx\jm^\m_\xx+ {1\over \sqrt{Z_K}}(\bar\h_\xx\psi_\xx
+\bps_\xx\h_\xx)+ \jm^\m_\xx A^\m_\xx\right]} \label{fi}
\end{equation}
where $\NN$ is a renormalization constant, $\h,A$ are external
fields, $\psi_\xx,\bps_\xx$ are Grassmann variables,
$\jm^\m_\xx=\bar\psi_\xx\g^\m\psi_\xx$ and $P_K(d\psi)$ is the
Grassmann integration with propagator
\begin{equation}
g_{\k,K}(\xx,\yy)=\int\!{\der^2\kk\over(2\p)^2 }\ \chi_{\k,K}(\kk)
{e^{-i\kk(\xx-\yy)}\over {\not\!\kk+\m_K}}\;,
\end{equation}
where the {\it smooth cutoff function} $\chi_{\k,K}(\kk)$ selects
the momenta $\k\le|\kk|\le K$, with $\k<1$, $K>1$. Finally $Z_K$
and $\m_K$ are the bare wave function renormalization and mass and
$\g^5$, $\g^\m$ are the Euclidean gamma matrices. The Schwinger
functions are defined as
\begin{equation}
S^{K,k}_{n,m}={\partial^{n+m}W_{\k,K}(A,\h)\over\partial
A^{\m_1}_{\zz_1}...\partial
A^{\m_m}_{\zz_m}\dpr\bar\h_{\xx_1}\dpr\h_{\yy_1}\ldots
\partial\bar\h_{\xx_n}\dpr\h_{\yy_n}}\Big|_{0}\;.
\end{equation}
In particular
\begin{equation}
G^\m_{\k,K}(\zz;\xx,\yy)={\partial^3 W_{\k,K}\over
\partial A^\m_\zz\dpr\bar\h_{\xx}\dpr\h_\yy
}\Big|_{0},\quad G_{\k,K}(\xx,\yy)={\partial^2 W_{\k,K}\over
\partial\bar\h_{\xx}\dpr\h_\yy
}\Big|_{0}
\end{equation}
and $G^\m_{\k,K}(\zz;\xx,\yy)=-i\e_{\m,\n}G^{\m,5}_{\k,K}(\zz;\xx,\yy)$.

The presence of the ultraviolet cutoff $K$ and the infrared cutoff
$\k$ makes the functional integral (\ref{fi}) well defined;  to
carry out the renormalization program at non-perturbative level we
have to prove that there exist $K$-depending bare parameters such
that, in the limit $\k\to 0, K\to\io$, the Schwinger functions
verify the Osterwalder-Schrader axioms, (OS), \cite{[OS]}. Our
basic result \cite{[BFM]} is the following theorem.
\begin{theorem}
Given $\l$ small enough and $\m\ge 0$, there exist bare parameters
\begin{equation}
Z_K=K^{-\h}\big(1+{\rm O}(\l^2)\big) \quad, \quad\m_K=\m
K^{-\bar\h}\big(1+{\rm O}(\l))\;,
\end{equation}
with $\h=a\l^2+O(\l^4)$ and $\bar\h=-b\l+o(\l^2)$, $a,b>0$, such
that $\lim_{k^{-1},K\to\io} S^{K,k}_{n,m}$ exist at coinciding
points and verify the Osterwalder-Schrader axioms.

In particular, in the massless case $\m= 0$,
\begin{equation}
\lim_{\k^{-1},K\to\io} G(\xx,\yy)_{\k,K}= (1+f_\l){\!\!
\not\!\xx-\!\!\not\!\yy\over |\xx-\yy|^{2+\h}}\;.
\end{equation}
with $f_\l=O(\l)$ indipendent from $\xx,\yy$.
\end{theorem}
The above theorem says that, by choosing properly the bare wave
function and mass renormalization, with a singular behavior as the
cutoffs are removed, one gets a set of finite Schwinger functions,
for which a QFT for the massive Thirring model is obtained via the
reconstruction theorem in \cite{[OS]}. The proof is based on the
new methods introduced in \cite{[BM]}, which allow us to overcome
the well known technical problem posed by the combination of a
non-perturbative setting based on multiscale analysis
\cite{[G],[P]} with the necessity of exploiting cancellations due
to the local symmetries. Such cancellations are established by
suitable WI valid at each scale and, contrary to the WI formally
valid when all the cutoffs are removed, they have corrections due
to the cutoffs. The crucial role of WI in the construction of the
theory is a feature that the Thirring model shares with realistic
models like $QED$ or the Electroweak theory in $d=4$, requiring WI
even to prove the perturbative renormalizability. Note that this
feature is absent in the models previously rigorously constructed
by functional integral methods, like the massive {\it Yukawa}
model \cite{[Le]} or the massive {\it Gross-Neveu} model
\cite{[GK]}. A delicate point in the proof in \cite{[BFM]} of the
above theorem is the verification of the positivity axiom, as the
cutoff we have chosen destroy the positivity definiteness and it
would be quite difficult to prove it directly after the cutoffs
are removed. We overcome this problem by considering a functional
integral with a {\it lattice regularization} containing a Wilson
term (to avoid a spurious singularity); in this case the
positivity property is automatically satisfied and one has to
choose the parameters $\l_a(\l)$, $Z_a(\l)$, $\m_a(\l)$, if $a$ is
the lattice step, in order to control the limit $a\to 0$. We have
then showed that it is possible to choose them so that
$\lim_{k^{-1},K\to\io} S^{K,k}_{n,m}=\lim_{a\to 0} S^{a}_{n,m}$,
hence positivity follows. Note finally that it has been claimed
that the functional integral (\ref{fi}) can be exactly computed
for massless fermions and no cutoffs, see for instance
\cite{[FGS]}, by introducing a boson field and using that that
fermionic determinants are {\it quadratic} if the Bose field is
regular. In this way one recovers the same results of the exact
solution, but mathematically this procedure is not justified (the
functional integral is over all the possible fields) and, as we
will see below, it leads to wrong conclusions.

\section{Anomaly Renormalization}
\begin{figure}[h]
\centering
\includegraphics[width=.3\textwidth,angle=0]{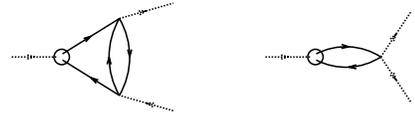}
\caption{\label{fig2} First and second order conribution to the anomaly;
The small circle represents $C_\o$}
\end{figure}

We shall only consider the WI in the massless case, $\m_K= 0$; the
presence of the cutoff function $\chi_{\k,K}$ breaks the
continuous symmetries, $\psi_\xx \to
e^{i\a_{\xx}+i\a^5_\xx\g^5}\psi_\xx$ and $\bps_\xx \to \bps_\xx
e^{-i\a_{\xx}+i\a^5_\xx\g^5}$, so that there are corrective terms
with respect to the formal WI (obtained formally neglecting all
cutoffs):
\begin{widetext}
\begin{eqnarray}
\pp^\m \widehat G^\m_{\k,K}(\pp;\kk)=
\widehat G_{\k,K}(\kk-\pp)-\widehat G_{\k,K}(\kk)
+
\int\!{\der^2\kk'\over (2\p)^2}\
C^\m_{\k,K}(\kk',\kk'-\pp)\la\bps_{\kk'}\g^\m\ps_{\kk'-\pp}\big|\ps_\kk\big|\bps_{\kk-\pp}\ra_{\k,K}
\label{gg}\\
\pp^\m \widehat G^{\m,5}_{\k,K}(\pp;\kk)=
\g^5\widehat G_{\k,K}(\kk-\pp)-\g^5\widehat G_{\k,K}(\kk)
+
\int\!{\der^2\kk'\over (2\p)^2}\
C^{\m}_{\k,K}(\kk',\kk'-\pp)\la\bps_{\kk'}\g^\m\g^5\ps_{\kk'-\pp}
\big|\ps_\kk\big|\bps_{\kk-\pp}\ra_{\k,K}\;;  \label{ggb}
\end{eqnarray}
the function $C^\m_{\k,k}(\kk_+,\kk_-)$ is given by
$\left(\chi^{-1}_{\k,K}(\kk_-)-1\right)\kk_-^\m-
\left(\chi^{-1}_{\k,K}(\kk_+)-1\right)\kk_+^\m$; $|$'s single out
the cluster of fields w.r.t. which the truncation of the
expectation is taken. The above expression can be perturbative
checked at lowest orders using the (trivial) identities
\begin{eqnarray}
\pp^\m\widehat g_{\k,K}(\kk-\pp)\g^\m\G\widehat g_{\k,K}(\kk)=
\G\big[\widehat g_{\k,K}(\kk-\pp)-\widehat g_{\k,K}(\kk)\big]
+C^{\m}_{\k,K}(\kk,\kk-\pp)\widehat g_{\k,K}(\kk-\pp)
\g^\m \G\widehat g_{\k,K}(\kk)\;,\label{iv}
\end{eqnarray}
for $\G=1,\g^5$.
\end{widetext}
The last addenda in the above WI involves the average of an higly
non-local and complex operator, but removing cutoffs, as
proven in \cite{[BFM]}, they can
be written in a remarkable simple form.
\begin{theorem}
In the same hypothesis of Theorem 1,
\begin{eqnarray}
&&\int\!{\der^2\kk'\over (2\p)^2}\
C^\m_{\k,K}(\kk',\kk'-\pp)\la\bps_{\kk'}\g^\m\ps_{\kk'-\pp}\big|\ps_\kk\big|\bps_{\kk-\pp}\ra_{\k,K}
\notag\\
&&=\a_+\pp^\m \widehat G^\m_{\k,K}(\pp;\kk)+H_{\k,K}(\pp;\kk)\label{jga}\\
&&\int\!{\der^2\kk'\over (2\p)^2}\
C^{\m}_{\k,K}(\kk',\kk'-\pp)\la\bps_{\kk'}\g^\m\g^5\ps_{\kk'-\pp}\big|\ps_\kk\big|\bps_{\kk-\pp}\ra_{\k,K}
\notag\\
&&=-\a_-\pp^\m \widehat G^{\m,5}_{\k,K}(\pp;\kk)+H^5_{\k,K}(\pp;\kk)\label{jg}
\end{eqnarray}
where $\a_+$ and $\a_-$ are suitable functions of $\l$ such that
\begin{equation}
\a_\pm={\l\over 2\pi} \pm  c_2\l^2+{\rm O}(\l^3)
\end{equation}
and
$c_2$ strictly negative; moreover,
for fixed non-zero $\kk,\pp$,
$\lim_{\k^{-1},K\to\io }H_{\k,K}(\pp;\kk)=
\lim_{\k^{-1},K\to\io }H^{5}_{\k,K}(\pp;\kk)=0$
\end{theorem}
Hence the WI we get in the functional integral approach have the
same form as those found in the exact approach, see (\ref{w1}) and
(\ref{w2}), but the anomaly coefficients, instead of by (5), are
given by $a^{-1}=1-{\l\over 2\pi}+c_+\l^2+O(\l^3)\virg \bar
a^{-1}=1+{\l\over 2\pi}+c_+\l^2+O(\l^3)$. The anomaly coefficients
are not linear in the bare coupling ({\it the anomaly
non-renormalization is violated }), contrary to (5); hence the
theory found in the massless limit starting from the functional
integral (6) has different properties with respect to the one
constructed by the exact solution. The presence of such anomaly
renormalization can be checked in standard perturbation theory, by
calculating the two graphs of Fig. \ref{fig2}, but the proof of
the non-perturbative bounds of Theorem 2 requires a careful
mathematical analysis, see \cite{[BFM]}.

\section{Additional anomalies}
The two point function $\widehat G_{k,K}(\kk)$ verifies the SD
equation
\begin{equation}
{\widehat G_{k,K}(\kk)\over \widehat g_{\k,K}(\kk)} = {1\over
Z_K}- \l \int {\der^2\pp\over (2\p)^2}\ \g^\m\widehat
G^\m_{k,K}(\pp;\kk-\pp)\;. \label{sd}
\end{equation}
Inserting the explicit expression
of $\widehat G^\m_{\k,K}$ obtained form the WI, we
obtain
\begin{eqnarray}
&&{\widehat G_{k,K}(\kk)\over\widehat g_{\k,K}(\kk)} ={1\over Z_K}
-\l(a-\bar a) \int\!{\der^2\pp\over (2\p)^2}
{\widehat G_{k,K}(\kk-\pp)\over \not\!\pp}\label{vvv}\\
&&-\int\!{\der^2\pp\over (2\p)^2} \left[{a
H_{\k,K}(\pp;\kk-\pp)\over \not\!\pp} +{\bar a\g^5
H^5_{\k,K}(\pp;\kk-\pp)\over \not\!\pp}\right]\;.\notag
\end{eqnarray}
{\it If} the last term in (\ref{vvv}) were vanishing in the limit
$\k^{-1}, K\to\io$, one would get a closed equation for $\widehat
G(\kk)$, which is identical to the closed equation obtained in
\cite{[J]}, that is (3). {\it This is not what happens}; as
already noticed, $H_{\k, K}(\pp;\kk)$ and $H^5_{\k, K}(\pp;\kk)$
are vanishing in the limit $\k^{-1}, K\to\io$ at $\kk,\pp$ fixed,
but {\it not} if $\pp$ is integrated. Intuitively this can be
understood by noting that the integral involves momenta close to
the u.v. cutoff scale $K$, where $H^{2,1}_{\k, K,\o,\o'}$ is not
small at all. In other words: {\it even if the WI and the SD
equation are true, in the limit $\k^{-1}, K\to\io$, the closed
equation obtained by combining the two identities is not verified;
this is a new anomaly} which is hard to see in a purely
perturbative approach and in fact it was never noticed before. One
could guess that the fact that the last term in (\ref{vvv}) is not
vanishing in the limit of removed cutoffs should imply that there
is no closed equation for $\widehat G(\kk)$. Instead, we proved in
\cite{[BFM]} another crucial identity.
\begin{theorem} In the same hypothesis
of previous theorems, the integral
in (\ref{vvv}) is equal to
\begin{eqnarray}
\s {1\over Z_K} + \r {\widehat G_{k,K}(\kk)\over\widehat
g_{\k,K}(\kk)}+ R_{\k,K}(\kk)\label{hhhg}
\end{eqnarray}
with O$(\l^2_K)$, non-zero $\s$ and $\r$; furthermore, for fixed
non-zero $\kk$, $\lim_{\k^{-1}, K\to\io}  R_{\k,K}(\kk)=0$.
\end{theorem}
By inserting
(\ref{hhhg}) in (\ref{vvv}) we get a closed equation which
is {\it different} from
the one assumed in \cite{[J]}; in particular, the relation
(\ref{xxx}) is replaced by
\begin{equation}
\h= {\l\over 2\p} {a-\bar a\over1 +\r}.
\end{equation}

\section{Conclusions}
We stress that one could construct a QFT corresponding to the
Thirring model also starting from a {\it non-local} interaction
$\int d\xx d\yy v(\xx-\yy) j^\m_\xx j^\m_\yy$ with $\hat
v(\kk)=e^{-\kk^2/ \L^2}$. If the cut-offs are removed in the order
$\L\to\io$, $K\to\io$, the results are the same as depicted it the
previous sections. On the contrary, if the cut-offs are removed in
the opposite order, $K\to\io$, $\L\to\io$, the results of the
exact solution are recovered. This means that the Adler-Bardeen
theorem is not in contrast with our result: if one assumes, as
in\cite{[AB]}, that the interaction is mediated by a boson field
and removes the fermionic u.v. cutoff before the bosonic one, then
the anomaly non-renormalization holds; in the opposite case new
features appear.

Finally, the combination of WI and SD equations is a rather
general technique; it is used, for instance, in QED in
\cite{[JBW],[GCB]} and in condensed matter physics in
\cite{[ILA]}. We have seen that such a method can really be
implemented in a full non-perturbative approach, but taking care
of unexpected anomalous features. It would be very interesting to
see if such features occur in 4-dimensional models.


{\it Acknowledgments} We are indebted with K. Gawedzki and G.
Gallavotti for enlightening discussions. P.F. gratefully
acknowledges the support of the Erwin Schr\"odinger Institute for
Mathematical Physics (Vienna).

\end{document}